\documentclass{aa}
\usepackage{graphicx}
\usepackage{rotate}
\usepackage{amsmath}
\usepackage{cancel}
\usepackage{amssymb}

\begin{document}

\title{The structure of sunspot penumbrae \\
IV. MHS equilibrium for penumbral flux tubes and the origin
of dark core penumbral filaments and penumbral grains}

\author{Borrero, J.M.\inst{1,2}}
\institute{High Altitude Observatory, NCAR\thanks{The National Corporation for Atmospheric 
Research (NCAR) is sponsored by the National Science Foundation.}, 3080 Center Green Dr, 
Boulder 80301, Colorado USA \and Max Planck Institut f\"ur Sonnensystemforschung, 
Max Plank Strasse 2, 37191 Katlenburg-Lindau, Germany}

\abstract{}
{We study the magnetohydrostatic equilibrium of magnetic flux tubes with circular cross sections embedded
in a magnetic surrounding atmosphere.}
{We solve the static momentum equation in 2.5D to obtain the thermodynamics that are consistent with
a prescribed velocity and magnetic fields.}
{We show that force balance is roughly satisfied if the 
flux tube's magnetic field is aligned with its axis. Equilibrium
is guaranteed if this magnetic field possesses a transverse component.
Several forms of this transverse field are investigated. 
The resulting magnetic field configurations are critically reviewed in terms of the results 
from spectropolarimetric observations. The thermodynamic structure 
that allows the flux tube to be in mechanical equilibrium is also calculated. 
We show that the inferred pressure, density and temperature
stratification reproduce intensity features similar to dark core penumbral filaments
and penumbral grains.}
{}
\titlerunning{MHS equilibrium of penumbral flux tubes}
\maketitle

%%% mathematical definitions
\def\nn{{\bf \nabla}}
\def\cro{\times}
\def\er{{\bf e_r}}
\def\et{{\bf e_{\theta}}}
\def\ex{{\bf e_x}}
\def\ez{{\bf e_z}}
\def\dep{(R,\theta)}
\def\depp{^{*}}
\def\deppp{(r,\theta)}
\def\hx{\mathcal{H}}
\def\gx{\mathcal{G}}
\def\mx{\mathcal{M}}
\def\nx{\mathcal{N}}

%%%%%%%%%%%%%%%%%%%%%%
\section{Introduction}%
%%%%%%%%%%%%%%%%%%%%%%

\noindent The filamentary structure of sunspot penumbrae is
 often explained in terms of flux tubes (Bellot Rubio 2003; Solanki 2003;
Thomas \& Weiss 2004). The equilibrium configuration 
for such penumbral flux tubes has been studied under the thin-flux tube 
approximation (Montesinos \& Thomas 1997; Schlichenmaier et al. 1998 and 
references therein). This approximation has the advantage that the governing 
equations become a set of one-dimensional equations. 
However, the thin-flux tube approximation has limited applicability 
in the solar photosphere, since the radius of the 
penumbral flux tubes is typically comparable to the pressure scale height. 

In this case, the problem becomes two-dimensional: density,
pressure and temperature vary over the cross section
of the flux tube. Spruit \& Scharmer (2006) and Scharmer \& Spruit (2006)
have suggested that, in this case, homogeneous flux tubes with a
circular cross section cannot be in mechanical equilibrium.
Such flux tubes are subject to stretching along the vertical
direction and flattening along the horizontal direction that changes
their shape and will likely destroy them. 

In this paper we investigate this conjecture and take a
first step towards a more realistic 
modeling of penumbral flux tubes beyond the thin-flux tube 
approximation. We find analytical solutions for the static 
momentum equation for horizontal flux 
tubes with a circular cross section, embedded in a surrounding 
atmosphere that harbors a potential magnetic field. 
The basic idea is to set up a generic magnetic and 
velocity field that satisfies certain conditions: Maxwell equations,
boundary conditions, observations etc. These magnetic field
and velocity vectors are brought into the momentum
equation. By assuming force balance, we are able to obtain
the density and pressure (and therefore temperature)
distribution inside the flux tube.

This paper is organized as follows: sections 2 and 3 are devoted to describing
 the basic equations and boundary conditions between the flux tube and the 
external atmosphere. In Section 4 we assume a particular form for the magnetic 
field in the surrounding atmosphere (potential) and, applying the boundary 
conditions, we study which configurations for the flux tube magnetic field are plausible.
A general form for the flux tube magnetic field is then obtained.
In section 5 we apply this general form for the flux tube magnetic field
to the static momentum equation, developing a treatment to obtain
the thermodynamic structure. Sections 6 presents 
particular cases for the flux tube magnetic field, and discusses: 
{\bf a)} the general appearance of the magnetic field vector 
and how it compares with results from spectropolarimetric observations;
{\bf b)} the thermodynamic structure inside the flux tube; {\bf c)}
solutions of the Radiative Transfer Equation in order to simulate
the observed continuum intensity emerging from such thermodynamic stratification.
In section 7 we critically review our results. Finally, section 8 summarizes
the main conclusions of our work.

%%%%%%%%%%%%%%%%%%%%%%%%%%%%%%%%%%%%%%%%%%%%%%%%%%
\section{Basic equations and boundary conditions}%
%%%%%%%%%%%%%%%%%%%%%%%%%%%%%%%%%%%%%%%%%%%%%%%%%%

\noindent Let us consider the case of a horizontal flux
tube that carries the Evershed flow and is embedded in a static plasma with
an inclined magnetic field. The flux tube's cross section is circular, with a 
radius $R$. For simplicity, its central position in the vertical direction will
be taken to be at $z=0$. The stationary momentum
equation can be written (in {\it cgs} units) as

\begin{eqnarray}
\rho ({\bf v} \nn) {\bf v} & = & -\nn P_g + \frac{1}{c} {\bf j} \cro {\bf B} + \rho {\bf g}
\end{eqnarray}

where $\rho$ and $P_g$ are the density and gas pressure. {\bf B}, {\bf j}, {\bf v} and {\bf g}
are the magnetic field, velocity, current and gravity vectors respectively. The gravity is taken
as ${\rm{\bf g}} =  - g \ez$ with $g = 2.74 \times 10^{4}$ cm s$^{-2}$.
Eq.~1 describes the force balance between the inertial force,
the pressure gradient, the Lorentz force and the gravity force. Let us use 
cylindrical coordinates with the axis of symmetry along the tube's axis
($x$-direction; radial direction along the penumbra).  According to Borrero et al. 
(2004, 2005 and 2006a; hereafter referred to as papers I,II and III), 
the properties of the flux tubes and the background atmosphere 
change rather smoothly along the radial direction in the
penumbra. These changes are quantitatively smaller 
than the variations across the plane perpendicular to the tube
axis, where large gradients can be present at very small scales. We
will therefore neglect the variation of any quantity along the $x$ coordinate.
This simplifies our problem to a large extent, since now we
consider only dependencies on the plane containing the cross 
section of the flux tube (plane $yz$; here we can also consider polar 
coordinates). We can write, separately for the flux tube
interior (index 't', tube) and exterior (index 's', surroundings)

\begin{equation}
{\bf B} = 
\begin{cases}
    {\bf B_s} = B_{rs}(r,\theta) \er +B_{\theta s}(r,\theta) \et +
    B_{xs}(r,\theta) \ex & r>R
    \\
    {\bf B_t} = B_{rt}(r,\theta) \er +B_{\theta t}(r,\theta) \et +
    B_{xt}(r,\theta) \ex & r<R
    \end{cases}
\end{equation}

\begin{equation}
{\bf v} =  
\begin{cases}
    {\bf v_s} = 0 & r>R 
    \\
    {\bf v_t} = v_{t0} \ex & r < R \;\;\;\;\;\; \rightarrow {\rm Evershed \;\;flow}\\
\end{cases}
\end{equation}

\noindent Note that although there is no dependence on the $x$-coordinate, the magnetic
field configuration is 3-dimensional. The normal component of the external and the 
internal magnetic field are assumed to vanish at the tube's boundary $r=R$. 
This is important in order to keep a clear distinction between the flux tube and the surrounding 
atmosphere:

\begin{eqnarray}
B_{rs}\depp = B_{rt}\depp = 0
\end{eqnarray}

\noindent where the superscript $^{*}$ denotes the location of the tube's boundary $\dep$. This 
boundary condition satisfies the continuity of the
normal component of the magnetic field vector across the
interface. Substituting our generic magnetic and velocity fields (Eqs.~2
and 3) into Eq.~1 we obtain,

\begin{equation}
\begin{cases}
    \er : 0 = -\frac{\partial P_g}{\partial r}+\frac{1}{c}(j_{\theta}
    B_x-j_x B_{\theta})-\rho g \sin \theta\\
    \et : 0 = -\frac{1}{r}\frac{\partial P_g}{\partial \theta}+\frac{1}{c}(j_x
    B_r-j_r B_x)-\rho g \cos \theta
    \end{cases}
\end{equation}

\noindent These equations are valid for both the flux tube's exterior and 
interior. $\theta=0$ lies along $y > 0$ and grows counter-clockwise. 
    We can make use of the relation $\nn \cro {\bf B} = \frac{4\pi}{c} {\bf j}$, 
    in order to obtain

\begin{eqnarray}
\frac{1}{r}\frac{\partial B_x}{\partial \theta} & = & \frac{4\pi}{c} j_r
\end{eqnarray}

%%%%%%%%%%%%%%%%%%%%%%%%%%%%%%%%%%%%%%%%%%%%%%%%%%%%%%%
\section{Pressure and density balance at the boundary}%
%%%%%%%%%%%%%%%%%%%%%%%%%%%%%%%%%%%%%%%%%%%%%%%%%%%%%%%

Both the radial component of external and internal magnetic
field vanish in the vicinity of the flux tube's boundary (Eq.~4). This leads to
continuity of the normal stress across the interface (Kippenhahn \&
M\"ollenhoff 1975). Therefore, by integrating the Maxwell stress tensor across
the flux tube's boundary we find total pressure balance (thermal plus magnetic)
between the flux tube's interior and its surroundings:

\begin{equation}
P_{gs} \depp + \frac{B_s^{2*}}{8\pi} = P_{gt} \depp + \frac{B_t^{2*}}{8\pi}
\end{equation}

\noindent Taking derivatives with respect to $\theta$ and regrouping terms, we obtain:

\begin{align}
\frac{\partial P_{gs}\depp}{\partial\theta}-\frac{\partial P_{gt}\depp}{\partial\theta}
& = \frac{1}{8\pi}\frac{\partial}{\partial\theta}
\left[B_{\theta t}^{2*}+B_{xt}^{2*}-B_{\theta s}^{2*}-B_{xs}^{2*}\right]
\end{align}

\noindent Terms in the left hand side of Eq.~8 can be written as a function of the
density using Eqs.~5 and 6,

\begin{align}
\frac{\partial P_g\depp}{\partial \theta} = & - \frac{j_r\depp B_x\depp R} {c}
-\rho\depp R g \cos\theta\notag \\ = & -\rho\depp R g \cos\theta - \frac{1}{8\pi}\frac{\partial B_x^{2*}}{\partial \theta}
\end{align}

\noindent Now, by substituting Eq.~9 into 8 we obtain

\begin{eqnarray}
\rho_t\depp-\rho_s\depp =  \frac{1}{8\pi R g \cos\theta} \frac{\partial
}{\partial \theta} \left\{B_{\theta t}^{2*}-B_{\theta s}^{2*}\right\} 
\end{eqnarray}

\noindent Note that Eqs.~7 and 10 link the gas pressure and density of the flux tube
and the surrounding atmosphere across the interface. They have been obtained
with only two assumptions: {\bf (a)} the radial component of the flux tube and
surrounding magnetic field has to vanish at the boundary $r=R$ and, {\bf (b)} velocity and
magnetic field vectors are constant in the direction of the tube's axis.

%%%%%%%%%%%%%%%%%%%%%%%%%%%%%%%%%%%
\section{Potential external field}%
%%%%%%%%%%%%%%%%%%%%%%%%%%%%%%%%%%%

Let us consider the case in which the external magnetic field
is potential: $\nabla^2 \Phi_s=0$ and ${\bf B_s}=-\nn \Phi_s$. This problem 
can be solved by variable separation. In addition, the boundary conditions 
given by Eq.~4 must be satisfied

\begin{equation}
B_{sr}\depp = 0
\end{equation}

\noindent Consider (see papers I,II; 
see also Bellot Rubio et al. 2004; Bello Gonz\'alez et al. 2005) that the external atmosphere 
is nearly at rest, and possesses a magnetic field that 
is somewhat inclined with  respect to the vertical $z$-axis. We therefore, 
assume that ${\bf B_s}$ has the following form far away from the tube

\begin{align}
\lim_{r \to \infty} {\bf B_s} & =  B_0 \cos\gamma_0\ez
      +B_0 \sin\gamma_0\ex = B_0 \cos\gamma_0\sin\theta\er \notag
      \\ & +B_0\cos\gamma_0\cos\theta\et+B_0\sin\gamma_0\ex
\end{align}

\noindent where $B_0$ and $\gamma_0$ are the strength and inclination
(with respect to the vertical to the solar surface) of the external
field far away from the flux tube. This yields the following 
solution for ${\bf B_s}\deppp$

\begin{eqnarray}
\begin{split}
{\bf B_s} & = B_0 \sin\gamma_0 \ex + B_0 \sin\theta\cos\gamma_0
  (1-\frac{R^2}{r^2})\er+\\ & B_0 \cos\theta\cos\gamma_0
  (1+\frac{R^2}{r^2})\et
\end{split}
\end{eqnarray}

%%%%%%%%%%%%%%%%%%%%%%%%%%%%%%%%%%%%
\subsection{Uniform Internal Field}%
%%%%%%%%%%%%%%%%%%%%%%%%%%%%%%%%%%%%

For the flux tube we now assume that both the magnetic field and
Evershed flow are constant and parallel to its axis: ${\bf B_t}=B_{t0}\ex$ and
${\bf v_t}=v_{t0}\ex$. By inserting ${\bf B_t}$
(as well as the external ${\bf B_s}$ given by Eq.~13) into Eq.~10  we obtain a relation 
between the external and internal densities:

\begin{eqnarray}
\rho_t\depp-\rho_s\depp = \frac{B_0^2\sin\theta\cos^2\gamma_0}{\pi R g}
\end{eqnarray}

Adopting typical values: $B_0 \simeq 1500$ G, $\gamma_0 \simeq 60^{\circ}$, 
$R \simeq 100$ km, we can estimate the density difference across the interface to be

\begin{eqnarray}
\rho_t\depp & \simeq & \rho_s\depp + 7 \times 10^{-7} \sin\theta \;\;\; {\rm [g cm^{-3}]}
\end{eqnarray}

\noindent The physical significance of this equation can be understood in terms
of the total pressure balance given by Eq.~7 (see also discussion in Borrero et al.
2006b). At the top and bottom of the flux tube, the transverse (azimuthal
and radial) component of the external field (Eq.~13) vanishes: $B_{\theta s}(\pm \pi/2,R) = 
B_{rs}(\pm \pi/2,R) = 0$. This creates a pressure imbalance at these points that 
makes the flux tube expand vertically ($z$-direction) and flatten
horizontally ($y$-direction). Eq.~14 shows that, in order to avoid this effect and keep the flux 
tube in mechanical equilibrium, the upper portions of the flux tube 
must be denser that the external
atmosphere: $\rho_t > \rho_s$ such that it becomes antibuoyant. The opposite happens
in the lower half of the flux tube: $\rho_t < \rho_s$. According to Eq.~15,
the density difference needed is about $7\times 10^{-7}$ g cm$^{-3}$ for typical penumbral
conditions.

Another way of looking at the problem is in terms of the pressure scale heights.
Considering total pressure balance and assuming that (both inside and 
outside the flux tube) the gas pressure decays exponentially, it is easy
to see that, for a homogeneous field inside the flux tube, the fact that the 
transverse component of the external field vanishes at the top and 
bottom of the flux tube, implies a larger pressure scale height inside the flux 
tube than outside in the lower half of the flux tube, but smaller 
than the external in the upper half.

Table~1 shows the parameters of an exponential fit
to the vertical density and gas pressure stratification 
($\rho(z) \sim \rho_0 \exp{-z/H_\rho}$, $p(z) \sim p_0 \exp{-z/H_p}$) 
for two atmospheric models representing the sunspot umbra (Collados et al. 
1994) and the quiet Sun (Borrero \& Bellot Rubio 2002).
Taking these two models as the limiting cases for the external
atmosphere in which the flux tube is embedded, we can take $\rho_0$ from Table 1
and make $\rho_0=\rho_s^{*}$. Note that near the umbra, where the density at 
the continuum level is about $\rho_s^{*} \sim 1 \times 10^{-6}$ g
cm$^{-3}$, the existence
of a flux tube with a uniform magnetic field is plausible.

Indeed, very similar solutions have been recently presented 
by Sch\"ussler \& V\"ogler (2006) in 3D radiative MHD simulations (see their Fig.~3). 
In those simulations the intrusion into the external field is caused
by a nearly field-free gap, whereas in this situation the flux tube is filled with a
horizontal (pointing radially outwards in the penumbra) magnetic field.
Note that both situations are equivalent in terms of the density
balance presented in Eq.~10.

\begin{table}
\caption[]{Parameters of an exponential fit to the vertical
stratification of density and gas pressure for two atmospheric
models representing the sunspot umbra and quiet Sun. $P_0$ and
$\rho_0$ refer to the gas pressure and density at $z=0$.}
\begin{center}
\tabcolsep .6em
\begin{tabular}{|lcccc|}
\hline
Model & $P_0$ & $H_p$ & $\rho_0$ & $H_\rho$\\
& [dyn cm$^{-2}$] & [km] & [g cm$^{-3}$] & [km] \\\hline
\hline
Umbra & $2.76\times 10^5$ & $84$ & $1.01\times 10^{-6}$ & $92$ \\
Quiet Sun & $1.25\times 10^5$ & $120$ & $3.18\times 10^{-7}$ & $130$\\
\hline
\end{tabular}
\end{center}
\end{table}

In the outer penumbra, the right-hand side term in Eq.~14 becomes smaller as the external
field weakens and becomes more inclined: $B_0 \simeq 1000$ G, 
$\gamma_0 \simeq 65^{\circ}$:

\begin{eqnarray}
\rho_t\depp & \simeq & \rho_s\depp + 2 \times 10^{-7} \sin\theta \;\;\; {\rm [g cm^{-3}]}
\end{eqnarray}

Here the external density at the continuum level resembles more closely that 
of the quiet Sun: $\rho_s^{*} \sim 3 \times 10^{-7}$ g cm$^{-3}$.
On the one hand, the density perturbation (Eqs.~15-16) is smaller
than the background density. This makes it plausible the existence of a
flux tube with a homogeneous magnetic field and circular cross 
section. On the other hand, those two values are very 
similar. This places the current model at the limit of its validity. 
There are however, additional effects that contribute to the equilibrium, 
such as the subadiabaticity of the external  atmosphere (Borrero et al. 2006b), 
but this only works after the flux tube has been already deformed. 
Since we are interested in keeping a circular cross section, we will investigate 
other possibilities. Those could work alone or in combination with 
a deformed tube embedded in a convectively stable atmosphere.

%%%%%%%%%%%%%%%%%%%%%%%%%%%%%%%%%%%
\subsection{Non-uniform internal Field}%
%%%%%%%%%%%%%%%%%%%%%%%%%%%%%%%%%%%

Inspection of Eq.~10 reveals that if the flux tube magnetic field
possesses an azimuthal component such that

\begin{eqnarray}
\begin{split}
B_{\theta t}\depp & = \sqrt{\alpha \mathcal{F}(\theta)+B_{\theta s}^{2*}} \\ &
= \sqrt{\alpha \mathcal{F}(\theta)+4 B_0^2 \cos^2\theta\cos^2\gamma_0} 
\end{split}
\end{eqnarray}

\noindent the density difference (Eq.~10) that follows is

\begin{eqnarray}
\rho_t\depp-\rho_s\depp = \frac{\alpha}{8 \pi R g}\frac{1}{\cos\theta}\frac{\rm
  d \mathcal{F}(\theta)}{\rm d \theta}
\end{eqnarray}

\noindent We therefore have the freedom to choose $\alpha$\footnote{Note that $\alpha$
has dimensions of Gauss$^2$, where $\alpha/8\pi$ is equivalent
to a magnetic pressure: dyn cm$^{-2}$} and $\mathcal{F}(\theta)$ in such a
way that the right hand side term of Eq.~18 is much smaller than the background
density. For instance: $\alpha \ll 8 \pi R g \rho_s \depp$ and $\mathcal{F}(\theta)=\sin\theta$. \\

Considering now $\nn {\bf B_t}=0$ we can write

\begin{eqnarray}
\frac{\partial (r B_{rt})}{\partial r} = -\frac{\partial B_{\theta t}}{\partial
\theta}
\end{eqnarray}

\noindent Assuming again separation of variables in the following form

\begin{eqnarray}
B_{rt}(r,\theta)=\hx(\theta) \mx(r) \\
B_{\theta t}(r,\theta)=\gx(\theta) \nx(r)
\end{eqnarray}

\noindent with the following boundary conditions, that are used in order to satisfy
Eq.~4 and Eq.~17

\begin{eqnarray}
\mx(0)=0 \;\;\;\;\;\;\;\;\;\; \mx(R)=0 \;\;\;\;\;\;\;\;\;\; \nx(0)=0\\
\nx(R)=1 \;\;\;\;\;\;\;\;\;\; \gx(\theta) = \sqrt{\alpha \mathcal{F}(\theta)
+4 B_0^2 \cos^2\theta\cos^2\gamma_0}
\end{eqnarray}

\noindent Using Eqs.~(19)-(23) and after some algebra, it can be shown that a
possible magnetic field configuration is given by

\begin{align}
B_{rt}(\theta,r) = & \frac{r}{2R^2}(R-r) \frac{1}{\sqrt{\alpha \mathcal{F}(\theta)
    + 4 B_0^2 \cos^2\theta\cos^2\gamma_0}} \times \notag \\ & \times \left[\alpha \frac{\rm d
    \mathcal{F}(\theta)}{\rm d \theta} - 8 B_0^2 \sin\theta\cos\theta\cos^2\gamma_0\right] \\
\notag\\
B_{\theta t}(\theta,r) = & \frac{r}{R^2}(3r-2R) \sqrt{\alpha
    \mathcal{F}(\theta)+4 B_0^2 \cos^2\theta\cos^2\gamma_0}\\
B_{x t} = & B_{xt0}
\end{align}

Note also that ${\bf v_t}$ is not necessarily parallel to ${\bf B_t}$. This implies that
the magnetic field will change in time according to the induction equation.
However, once we consider the magnetic and kinematic properties of our model
(Eq.~3 and 24 through 26), it turns out that only the diffusive term of the induction
equation survives, resulting in magnetic field changes over 
very long time scales ($\sim$ days), which does not affect the discussion presented here.

The only restrictions in the choice of $\alpha$ and $\mathcal{F}(\theta)$ 
are: {\bf a)} the flux tube transverse magnetic field (given by Eq.~24 and 25) 
must not become singular; {\bf b)} the total transverse field must be much 
smaller than the field along the tube's axis: $\sqrt{B_{\theta t}^2+B_{r t}^2}<<B_{xt}$ 
(in order to be consistent with spectropolarimetric observations ; see Bellot Rubio et al. 2004)
; {\bf c)} $\rho_t^{*}$ (Eq.~18) must not become negative (see Sect.~4.1).

%%%%%%%%%%%%%%%%%%%%%%%%%%%%%%%%%%
\section{Thermodynamic structure}%
%%%%%%%%%%%%%%%%%%%%%%%%%%%%%%%%%%

Once the magnetic field in our 2-D domain is known (Eq.~13 for the external field
 and Eqs.~(24)-(26) for the flux tube field) we can apply the static momentum equation 
(Eq.~1; with a velocity field described by Eq.~3) in order to calculate a density 
and pressure distribution. For convenience we will work hereafter in Cartesian coordinates.

Since the external field is potential, ${\bf j} \cro {\bf B}=0$, Eq.~1 translates into

\begin{eqnarray}
\frac{\partial P_{gs}(z)}{\partial y} & = & 0 \\
\rho_s(z) & = & - \frac{1}{g} \frac{\partial P_{gs}(z)}{\partial z}
\end{eqnarray}

Therefore it is sufficient to prescribe the behavior of the
thermodynamic quantities far away from the flux tube $y \rightarrow \infty$. 
In our case we have adopted a temperature $T_s(z)$, density $\rho_s(z)$ and gas pressure 
stratification $P_{gs}(z)$ as given by the cool umbral model of Collados et al. (1994). 
In order to have a better coverage in deep layers this model has been extrapolated downwards
up to 1 Mm with typical solar stratification resulting from 3D numerical simulations
(see Cheung 2006; Fig.~3.1 and 3.3).

For the internal (flux tube) atmosphere we must consider that the magnetic field
configuration given by Eqs.~(24)-(26) is no longer potential. In particular there is
a current oriented along the flux tube axis:

\begin{equation}
{\bf j_{t}} = j_{xt}\ex = \frac{c}{4\pi r}\left[\frac{\partial}{\partial r}(r B_{\theta t})
-\frac{\partial B_{rt}}{\partial \theta}\right] \ex
\end{equation}

In this case, considering the horizontal ($y$-axis) direction of static momentum equation (Eq.~5)
yields the pressure distribution inside the flux tube:

\begin{equation}
P_{gt}(y,z) = - \frac{1}{c} \int j_{xt} B_{zt} dy +C(z)
\end{equation}

\noindent where $B_{zt}$ is obtained from Eqs.~24-25 and $j_{xt}$ from Eq.~29. The integration
constant $C(z)$ can be evaluated by imposing total pressure balance at the flux tube boundary.

\begin{equation}
P_{gt}(\pm\sqrt{R^2-z^2},z) = P_{gs}(z)+\frac{B_0^2\sin^2\gamma_0-B_{xt0}^2-\alpha\mathcal{F}(\theta)}{8\pi}
\end{equation}

Once the pressure is known for every point inside the flux tube, the vertical component of 
the momentum equation yields the density:

\begin{equation}
\rho_t(y,z) = \frac{1}{g} \left[\frac{j_{xt}B_{yt}}{c}-\frac{\partial P_{gt}}{\partial z}\right]
\end{equation}

Given the analytical solution for the magnetic field, Eqs.~(24)-(26), the gas pressure and density also have 
analytical expressions that can be evaluated through Eqs.~(30)-(32). These solutions ensure
that there is no net force acting on the flux tube and it is
therefore in mechanical equilibrium. Note that the term involving the velocity field (advection term) in
Eq.~1 vanishes as long as we assume that the velocity is
concentrated inside the flux tube and that it is 
parallel to its axis ($x$-axis; Eq.~3). Once we have two-dimensional ($yz$) maps of the density and gas pressure,
the temperature can be evaluated using the ideal gas equation. A variable
mean molecular weight is used in order to account for the partial ionization of 
the different atomic species.

%%%%%%%%%%%%%%%%%%%%%%%%%%%%%%%%%
\section{Results for $\alpha=0$}%
%%%%%%%%%%%%%%%%%%%%%%%%%%%%%%%%%

%%%%%%%%%%%%%%%%%%%%%%%%%%
\subsection{Magnetic field}%
%%%%%%%%%%%%%%%%%%%%%%%%%%

The first example we will consider is $\alpha=0$. This case 
is particularly interesting because it is valid for any $\mathcal{F}(\theta)$. 
In this case, the magnetic field is as follows:

\begin{eqnarray}
\begin{split}
{\bf B_s}  = & B_0 \sin\gamma_0 \ex + B_0 \sin\theta\cos\gamma_0
  (1-\frac{R^2}{r^2})\er+\\ & B_0 \cos\theta\cos\gamma_0
  (1+\frac{R^2}{r^2})\et
\end{split}
\end{eqnarray}

\begin{eqnarray}
\begin{split}
{\bf B_t}  = & B_{xt0} \ex - \frac{2r}{R^2}(R-r) B_0 \sin \theta \cos \gamma_0 \er \\
& +\frac{2r}{R^2}(3r-2R) B_0 \cos\theta\cos\gamma_0\et
\end{split}
\end{eqnarray}

Fig.~1 shows the total magnetic field (top panel) and inclination with respect to the vertical (middle
panel) resulting from Eq.~33 and Eq.~34 with the following parameters: $B_0 = 1700$ Gauss, 
$\gamma_0 = 65^{\circ}$, $B_{xt0} = 1200$ Gauss, $R = 125$ km. The inclination with respect to the
vertical ($z$-axis) direction is defined as: $\gamma = \cos^{-1}(B_{z}/B)$. Bottom panel in Fig.~1
displays the field lines for the transverse field ($yz$ plane). 
Inside the flux tube, the magnetic field is mainly align with the $x$-axis: 
$B_{xt} >> \sqrt{B_{\theta t}^2+B_{rt}^2}$, but is pointing slightly upwards along the sides 
($\gamma \sim 80^{\circ}$) and downwards in its center: $\gamma \sim 100^{\circ}$.

In spite of the simplicity of this model, the overall configuration displays many 
similarities with recent investigations of the penumbral fine
structure. For instance: {\bf a}: at a height corresponding to the axis of the flux tube, 
the internal (flux tube) magnetic field is weaker than the external
(surrounding field) by an average amount of $600$ Gauss (papers I,II,III; see also 
Bellot Rubio et al. 2004); {\bf b}: the average flux tube magnetic field is 
mostly horizontal $\gamma \sim 90^{\circ}$, while the external field is 
inclined (next to the flux tube) about 50$^{\circ}$ with respect to the vertical; 
{\bf c}: an observer at disk center could see an
inclination for the magnetic field inside the flux tube larger than 90$^{\circ}$ despite the fact
that the flux tube axis is perpendicular to the observer. This is due to the fact that 
the transverse field inside the flux tube has a negative component $B_z < 0$ in its 
inner part (see Fig.1 bottom panel). The existence of magnetic flux returning into
the solar photosphere has appeared in many different works (see, e.g. Westendorp Plaza
et al. 1997; del Toro Iniesta et al. 2001; Schlichenmaier et al. 2004). In the present 
model, returning flux can be explained by either a flux tube pointing downwards or, even if its axis is 
aligned with the $x$-axis (as in Fig.~1), as a consequence of the transverse 
field lines: $B_{rt}$ and $B_{\theta t}$.

\begin{figure}
\begin{center}
\includegraphics[angle=0,width=8.5cm]{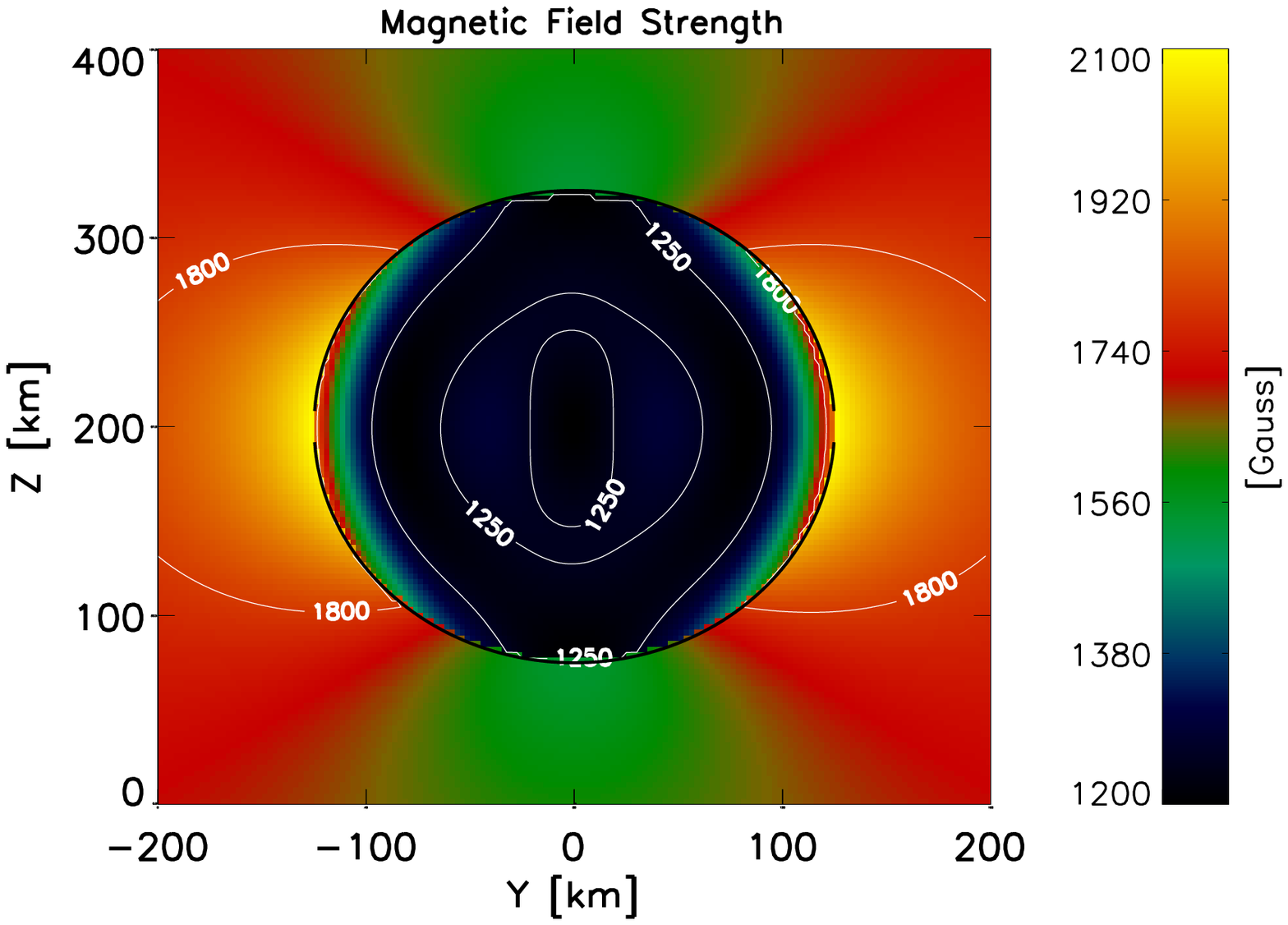} \\
\includegraphics[angle=0,width=8.5cm]{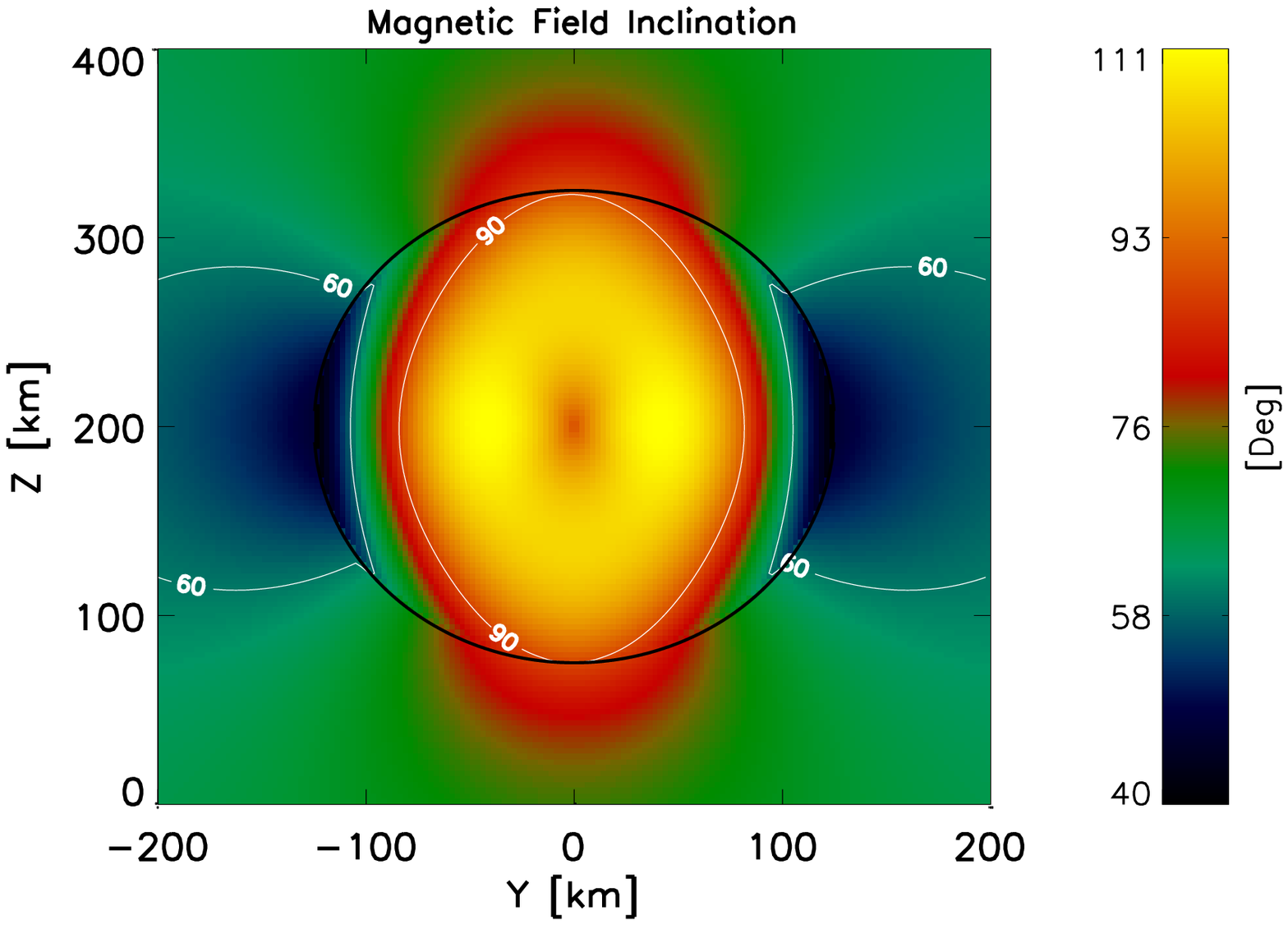} \\
\includegraphics[angle=0,width=8.5cm]{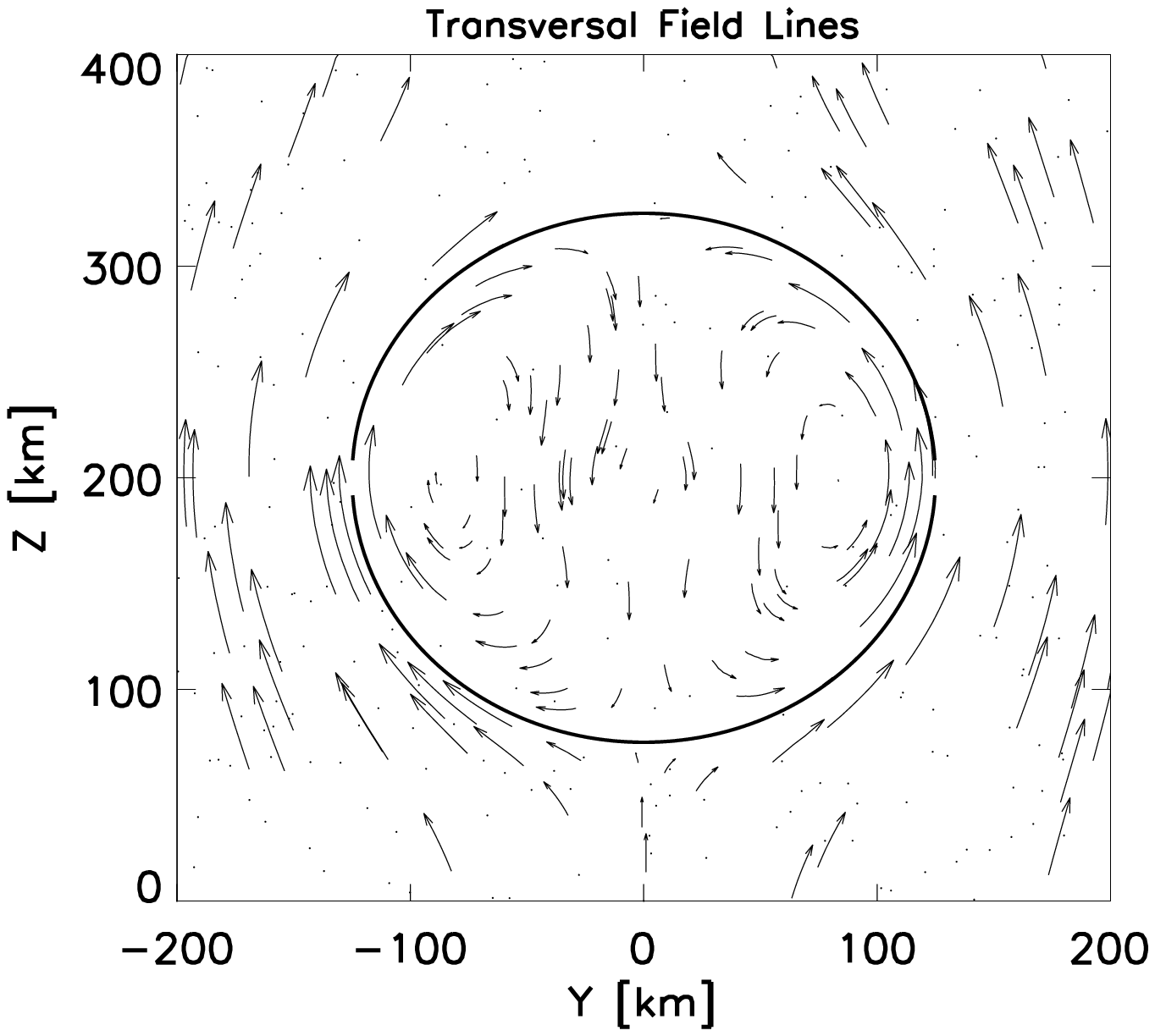}
\end{center}
\caption{Color plots of the total magnetic field strength (top panel) 
and magnetic field inclination (middle) for the following
parameters: $B_0 = 1700$ Gauss, $\gamma_0 = 65^{\circ}$, $B_{xt0} = 1200$ Gauss,
$R = 125$ km. Note that the strength of the magnetic field inside the flux tube along its axis
is $B_{xt0} = 1200$ Gauss, therefore almost all the magnetic field seen
inside the flux tube goes along the $x$ axis. Bottom panel displays the direction of 
the transverse field.}
\end{figure}

%%%%%%%%%%%%%%%%%%%%%%%%%%%%%%%%%%%
\subsection{Thermodynamic structure}%
%%%%%%%%%%%%%%%%%%%%%%%%%%%%%%%%%%%

Fig.~2 (top and middle panels) shows surface plots of the density and gas pressure
excess (with respect to the surrounding at each height) 
as obtained from the same magnetic field configuration as in Fig.~1.
The vertical scale $z$ has been shifted so that the the continuum $\tau_5=1$ level (denoted by
the middle white dashed line) in the surrounding atmosphere lies at $z=0$.
The center of the flux tube is located at $z_0 = 0$ km (same height as the 
continuum level in the surrounding atmosphere).

As in Sect.~4.1, the upper part of the flux tube is more dense 
that the surroundings at the same height, while the opposite happens in the lower 
part of the flux tube. This effect follows from the requirement of total pressure balance
at the top and bottom of the flux tube. In addition, according to
Eq.~18, there is no density change across the boundary 
between the flux tube and the magnetic surrounding: 
$\rho_s\depp=\rho_t\depp$.

\begin{figure}
\begin{center}
\includegraphics[angle=0,width=7cm]{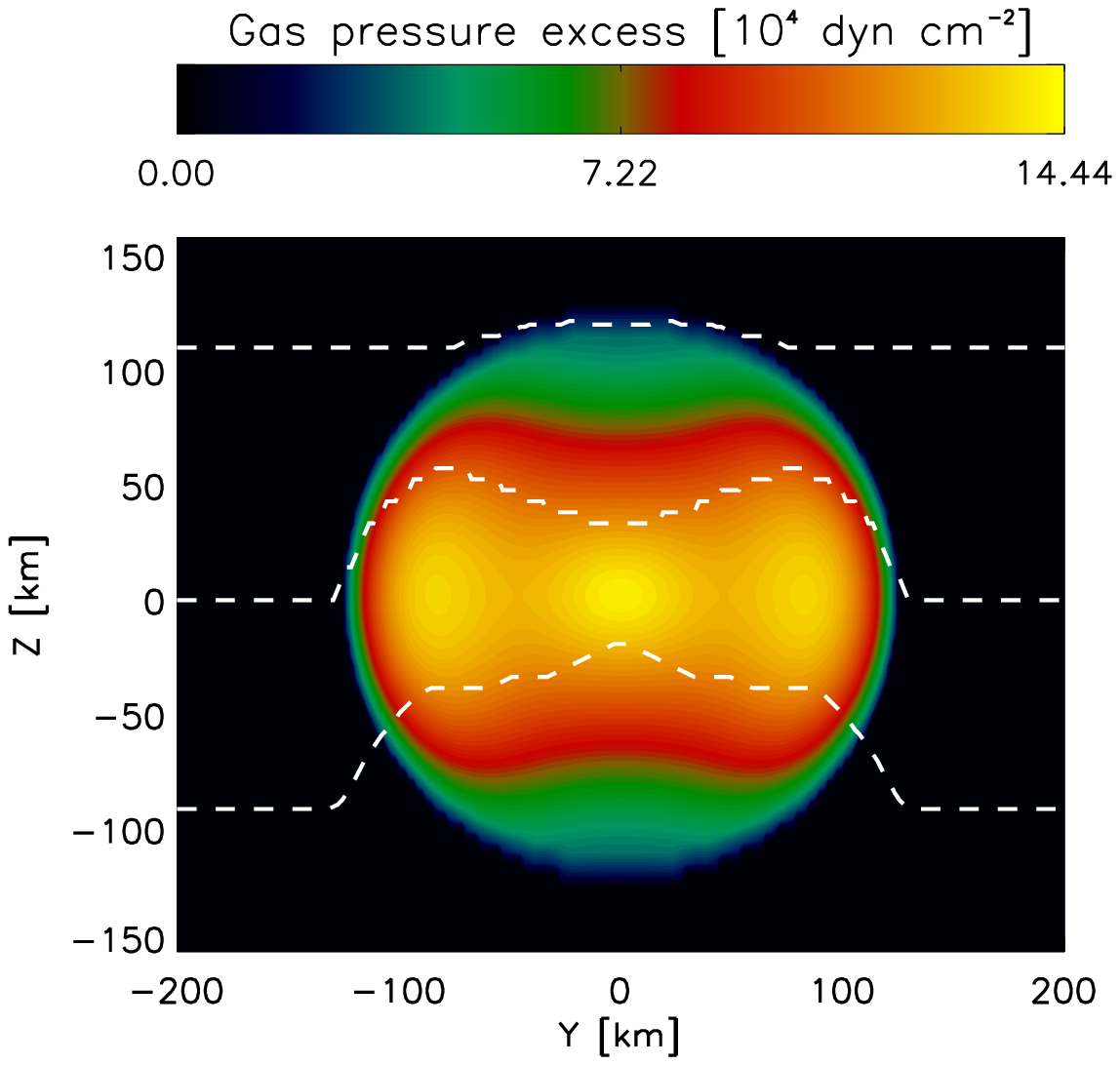} \\
\includegraphics[angle=0,width=7cm]{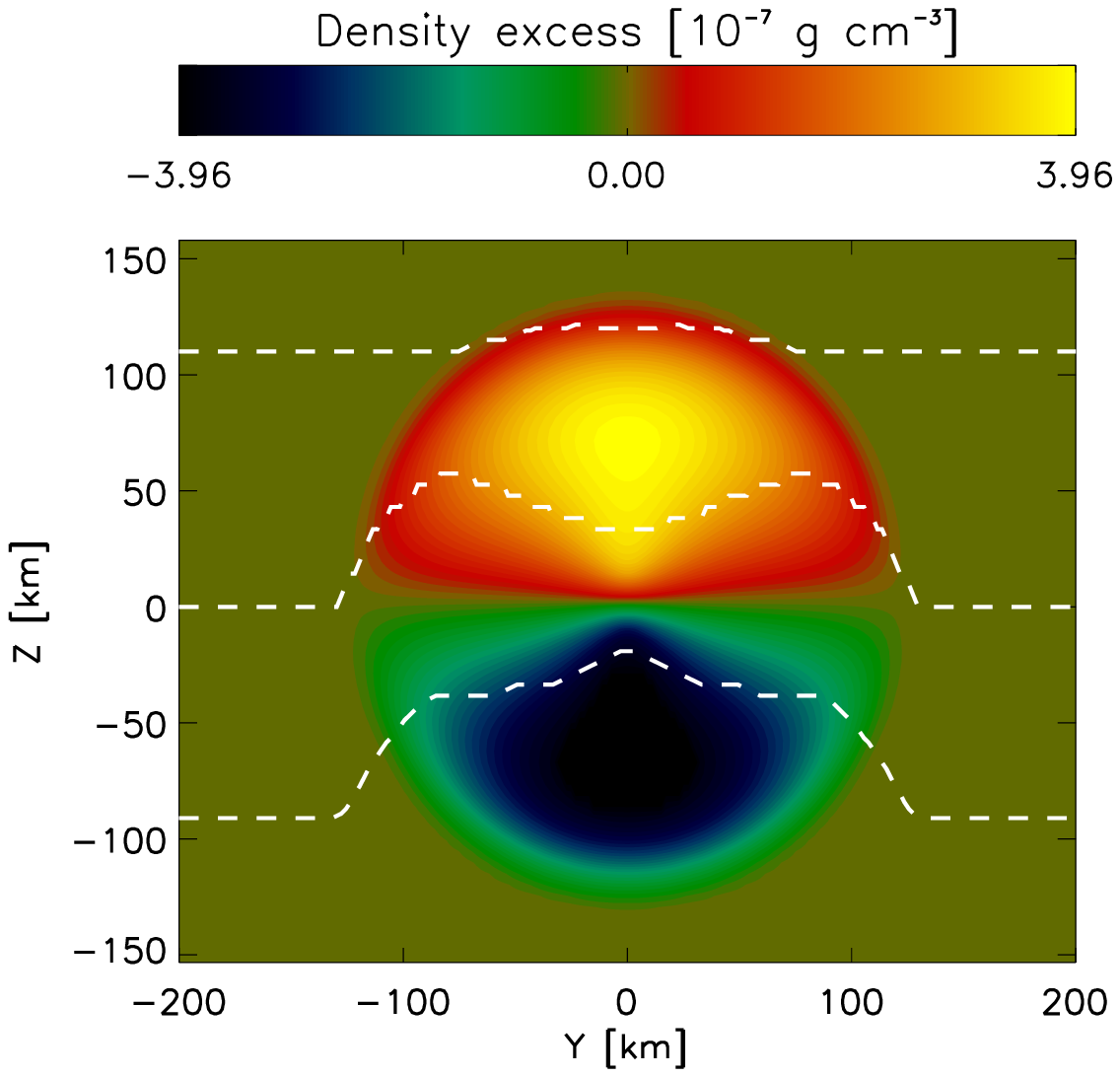} \\
\includegraphics[angle=0,width=7cm]{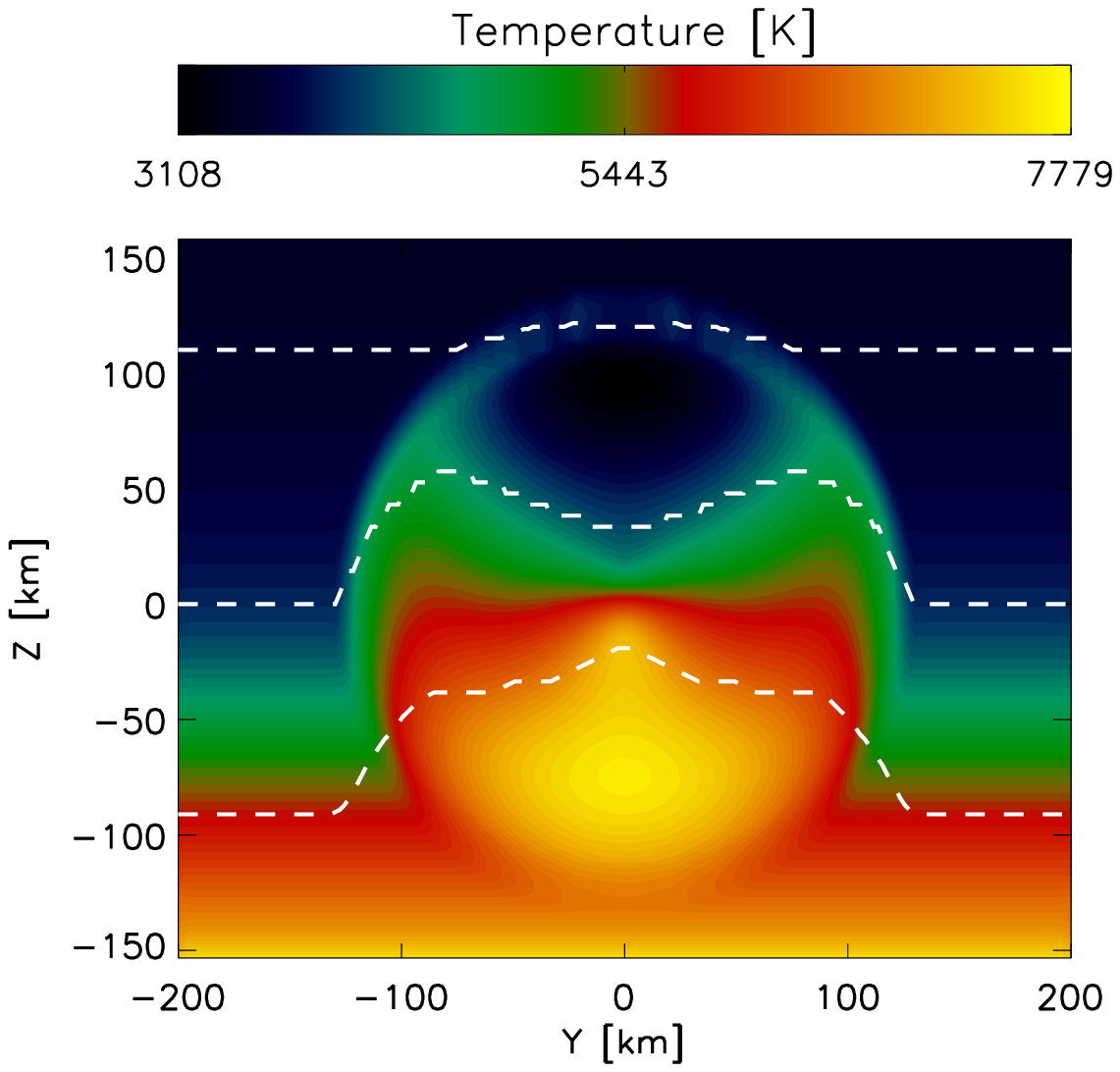}
\end{center}
\caption{{\it Top panel}: Gas pressure excess $P_g(y,z)-P_{gs}(\infty,z)$.
{\it Middle panel}: Same for density excess $\rho_g(y,z)-\rho_{gs}(\infty,z)$.
{\it Bottom panel}: Temperature distribution $T(y,z)$. Note that 
$P_{gs}(\infty,z)$ and $\rho_{gs}(\infty,z)$ decay exponentially with height $z$.
This distribution ensures that the flux tube is in mechanical equilibrium. Dashed 
white lines indicate (from lowermost to uppermost) the $\tau_5=10,1,0.1$ levels.}
\end{figure}

The temperature distribution is presented in Fig.~2 (bottom panel).
Note that the combination of low density and high gas 
pressure in the lower portions of the flux tube yields
large temperatures in this region. In addition, the high densities found 
in the upper half of the flux tube result in low temperatures there.
Depending upon the choice of our model parameters $B_0$, $\gamma_0$, 
$B_{xt0}$, $z_0$ and $R$, sometimes unrealistically high/low temperatures
(in the lower/upper half) may result. The evolution of any hot plasma element, that
rises adiabatically with typical entropy values corresponding to
the bottom of the convection zone, limits the temperature at
the photosphere to roughly $T_{\rm max} < 13500$ K. 
Although we are not aware of a lower limit for the temperature, 
we will also consider that any value below 2000 K is not realistic.

Fig.~3 presents a parameter study aiming at identifying
the regions where the resulting temperatures are within the
allowed range ($T_{\rm max} < 13500$ and $T_{\rm min} > 2000$) K.
Since we have 5 model parameters, we have kept 
three of them fixed while varying the other two:
$B_0$ and $B_{xt0}$ (top panel), $R$ and $\gamma_0$
(middle panel) and finally $R$ and $z_0$ (bottom panel).
In these plots, the solid lines separates the regions where
$T_{\rm max}$ and $T_{\rm min}$ are below/above 13500 K and 2000 K,
respectively. The shaded areas are the regions where both constraints
are met. Fig.~3 shows that, for very 
strong fields ($B_0 > 1800$ and $B_{xt0} > 1600$ G), it is difficult
 to satisfy the constraint of $T_{\rm max} < 13500$ K. The same happens for 
very vertical external magnetic fields ($\gamma_0 < 60^{\circ}$) and for 
thin flux tubes ($R < 50$ km). However, this does not mean that the model
presented here is not valid near the umbra. Indeed, Fig.~1 shows a magnetic
field configuration that very closely resembles that region.

We have also repeated the calculation of Fig.~3 but using as a background atmosphere
the mean penumbral model of del Toro Iniesta et al. (1994). This model
is similar (in its density and pressure stratification) to a quiet Sun model.
Our results indicate that temperatures below 13500 K are obtained
for $B_0$ and $B_{xt0} \lesssim 1200$ Gauss and R $>100$ km. This is consistent
with our discussion in Sect.~4.1.

\begin{figure}
\begin{center}
\includegraphics[angle=0,width=8cm]{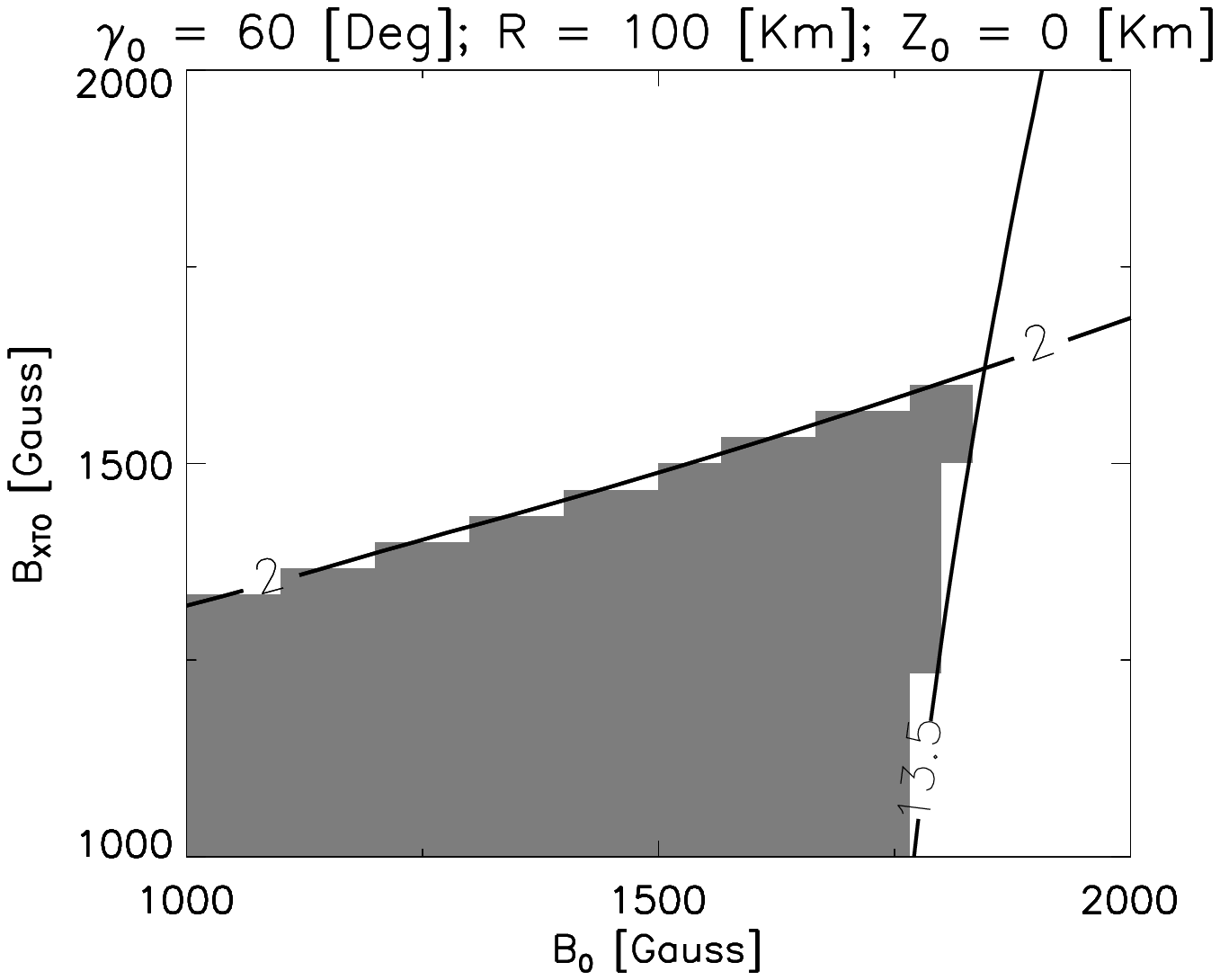} \\
\includegraphics[angle=0,width=8cm]{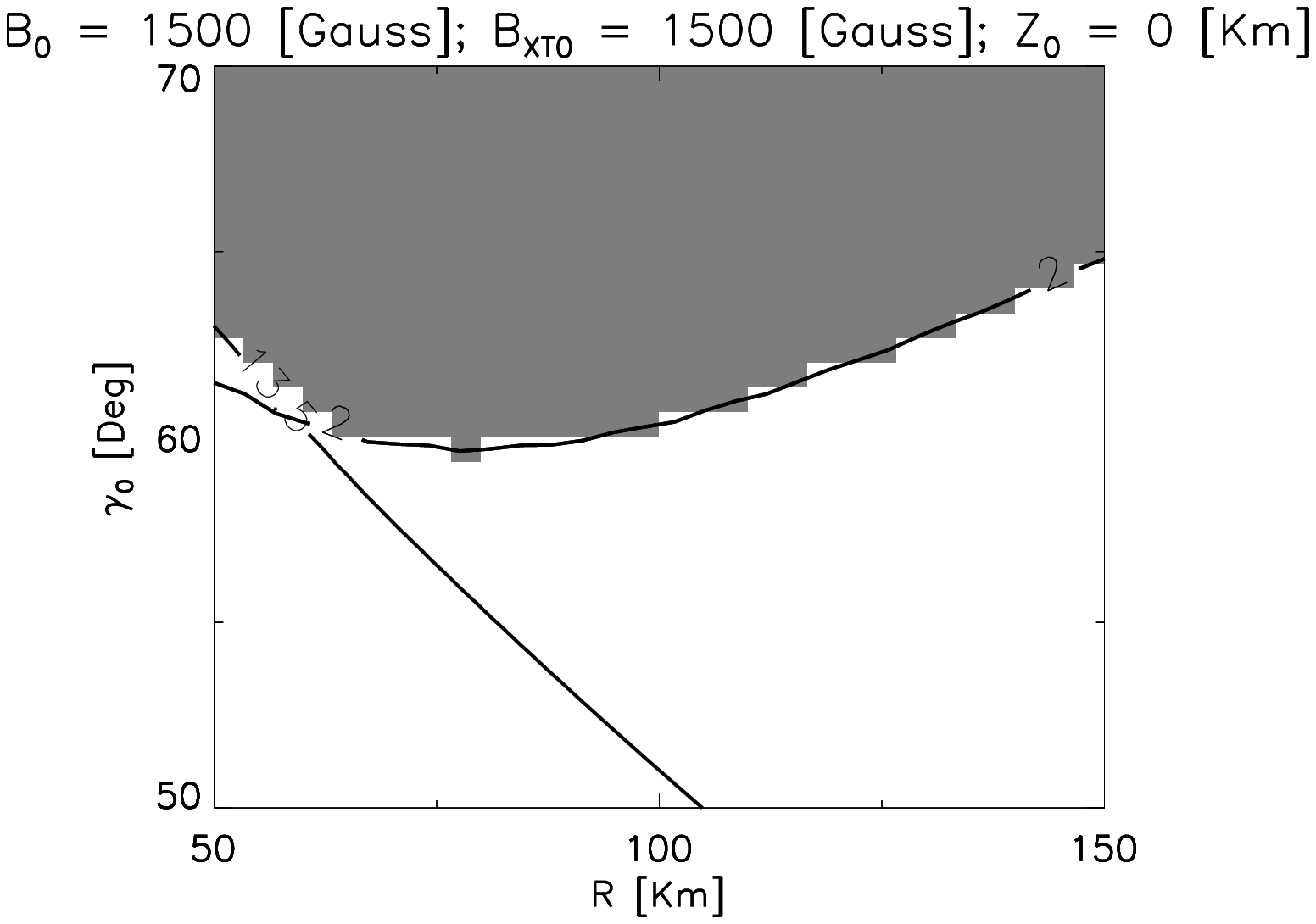} \\
\includegraphics[angle=0,width=8cm]{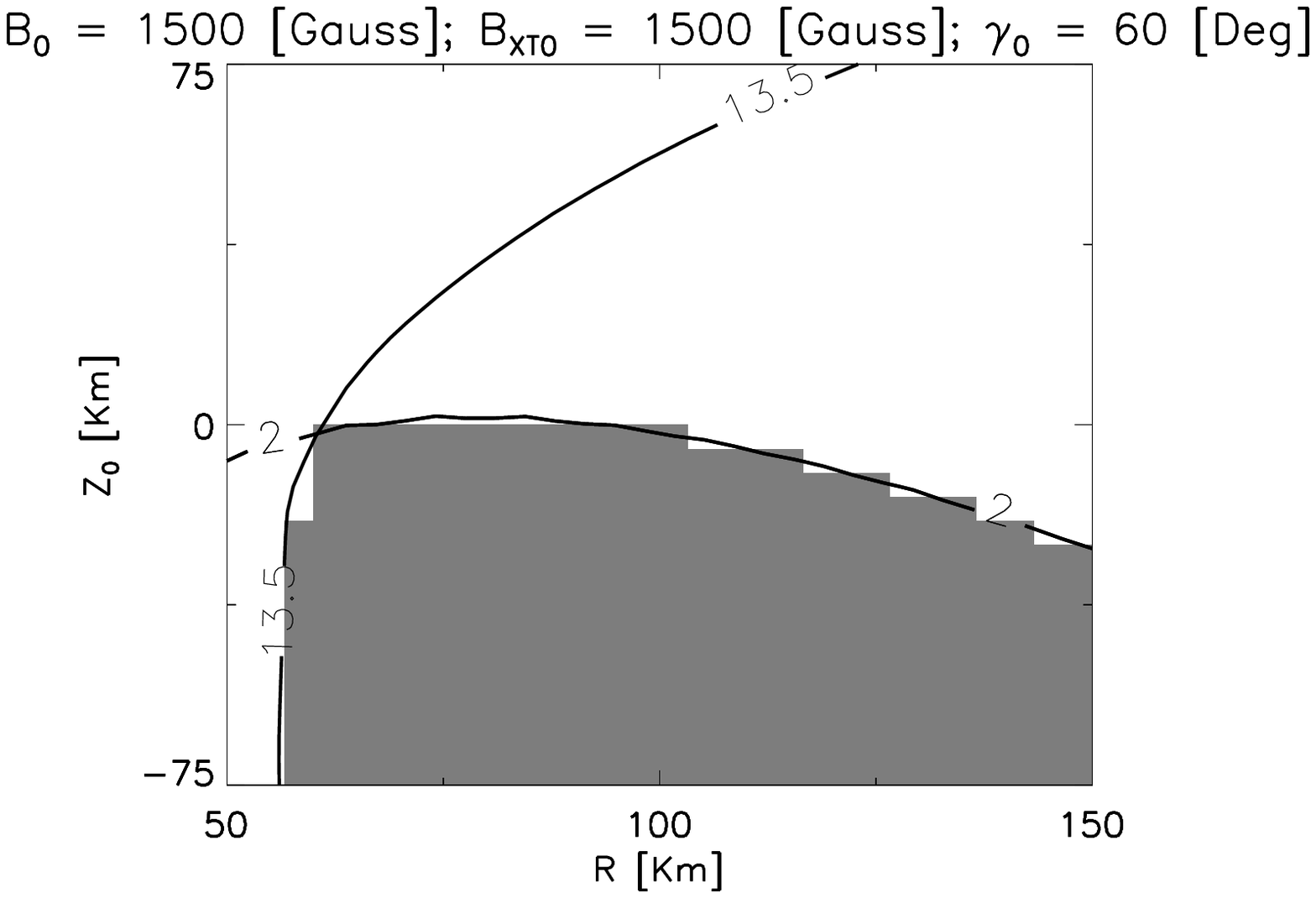}
\end{center}
\caption{Shaded areas indicate the range for which the model parameters
results in minimum temperatures (inside the flux tube) larger than 2000 K and maximum 
temperatures smaller than 13500 K. Only two parameters are allowed to
change at a time: $B_{xt0}$ and $B_0$ (top panels), $\gamma_0$ and $R$ (middle panel), 
$z_0$ and $R$ (bottom panel).}
\end{figure}

%%%%%%%%%%%%%%%%%%%%%%%%%%%%%%%%%%%%%%%%%%%%%%%%%%%%%%%%%%%%%%%%%%%%%%%%%%%%%%%%%%%%
\subsection{Emergent intensity: dark core penumbral filaments and penumbral grains}%
%%%%%%%%%%%%%%%%%%%%%%%%%%%%%%%%%%%%%%%%%%%%%%%%%%%%%%%%%%%%%%%%%%%%%%%%%%%%%%%%%%%%

High resolution observations of individual penumbral filaments in the G-band and 
continuum intensity at 4877 \AA~ (Scharmer et al. 2002; Rouppe van der Voort 
et al. 2004; S\"utterlin et al. 2004) show intensity enhancements 
at the edges of the filament. This is accompanied by a drop in 
the core intensity (Fig.~2 and 3 in Scharmer et al. 2002). 
The contrast varies from image to image but in general they are larger in G-band 
images than at 4877 \AA~. Ever since this discovery, this particular feature has been 
referred to as {\it dark core penumbral filaments}. Other penumbral features reveal 
{\it penumbral grains} composed of what looks like three different kernels next to 
each other (Fig.~6 in Rouppe van der Voort et al. 2004).

We have employed a numerical code to solve the Radiative transfer
equation (Frutiger 2000) and compute the values of the observed continuum 
intensity (at a reference wavelength of 5000 \AA) for each ray 
path along the $z$-direction. This returns the continuum intensity as
a function of the position $y$ across the filament. The result is convolved
with an Airy function with a FWHM of 100 km in order to account for the limited
spatial resolution of the observations.

We have made three different experiments in which continuum images at a
reference wavelength of 5000 \AA~ are produced by varying, one at a time,
different model parameters. Results are presented in Fig.~4: varying $z_0$ 
(top panel), $R$ (middle panel), $\gamma_0$ (bottom panel). The values for 
the other parameters are indicated in the  corresponding figures.

In the first experiment (Fig.~4; top panel) the only parameter that changes
is $z_0$ (central position of the flux tube with respect to the $\tau=1$
level of the background atmosphere). The intensity pattern that results
is very similar to a {\it dark core penumbral filament}. The intensity contrast
between background and bright feature is larger the higher the flux tube is 
located. Note that the background intensity is about 0.12 units of the quiet Sun 
intensity, resulting from the use of the cool umbral model by Collados et al. (1994).

For the second test (Fig.~4; middle panel) we fix $z_0=0$ while 
the flux tube radius $R$ changes from 75 to 150 km. For small 
flux tube radii we obtain an intensity pattern that can be identified with 
a {\it penumbral grain}. With increasing diameter the
emergent intensity morphology shifts to that of a {\it dark core filament}.
In the third test we explore the effects of a varying inclination
for the external field: $\gamma_0$ (see Fig.~4; bottom panel).
In this example, more vertical external fields yield {\it penumbral grains}. 
When $\gamma_0$ increases, the intensity at the center of the flux tube 
decreases and we start to see a {\it dark core filament}.
These are just examples of the kind of intensity features that the model is able
to produce. In all examples we have been careful to select only a set of
parameters that keep the minimum/maximum temperatures inside the flux tube
above/below 2000/13500 K, respectively.

Inversions of spectropolarimetric observations (papers I,II,III; 
Bellot Rubio et al. 2004; R\"uedi et al. 1998, 1999)
indicate that flux tubes are located high above the continuum
formation height in the inner
penumbra but they are found near to it in the outer penumbra
\footnote{A common interpretation of this result is that the flux tube
is located at a fairly constant height. However, the height separation
between the flux tube and the external $\tau=1$ level decreases away from the umbra
as the Wilson depression is reduced.}.  In addition, the external field becomes 
more horizontal as we move away from the umbra. 
Furthermore (and although not entirely clear based on
the interpretation of the Net Circular Polarization), it appears that
the radius of the flux tube increases radially in the penumbra\footnote{Magnetic
flux conservation, together with a decreasing magnetic field strength inside the flux 
tube as a function of $x$, leads to a radially increasing flux tube radius.} 
($x$-direction). To test what kind of radial feature would be obtained under these assumptions,
we have produced 50 different continuum images where the flux tube is initially
(inner penumbra) located at $z_0 = 50$ but descends linearly down to $z_0 = 0$ in 
the last image (further from the umbra). At the same time, the inclination of the external field changes linearly
from 60$^{\circ}$ in the first image to 70$^{\circ}$ in the last one. The radius, $R$,
was kept constant at 125 km, however. Results are presented in Fig.~5. Similarities
with observed features in high-resolution continuum images (see, e.g. Fig.~5
in Rouppe van der Voort et al. 2004) are remarkable.

Note that for this test we have assumed that some physical parameters
depend on the $x$ coordinate. This has no particular consequence on our
main assumptions as long as the changes along the radial direction
in the penumbra are small (see Sect.~2).

\begin{figure}
\begin{center}
\includegraphics[angle=0,width=7cm]{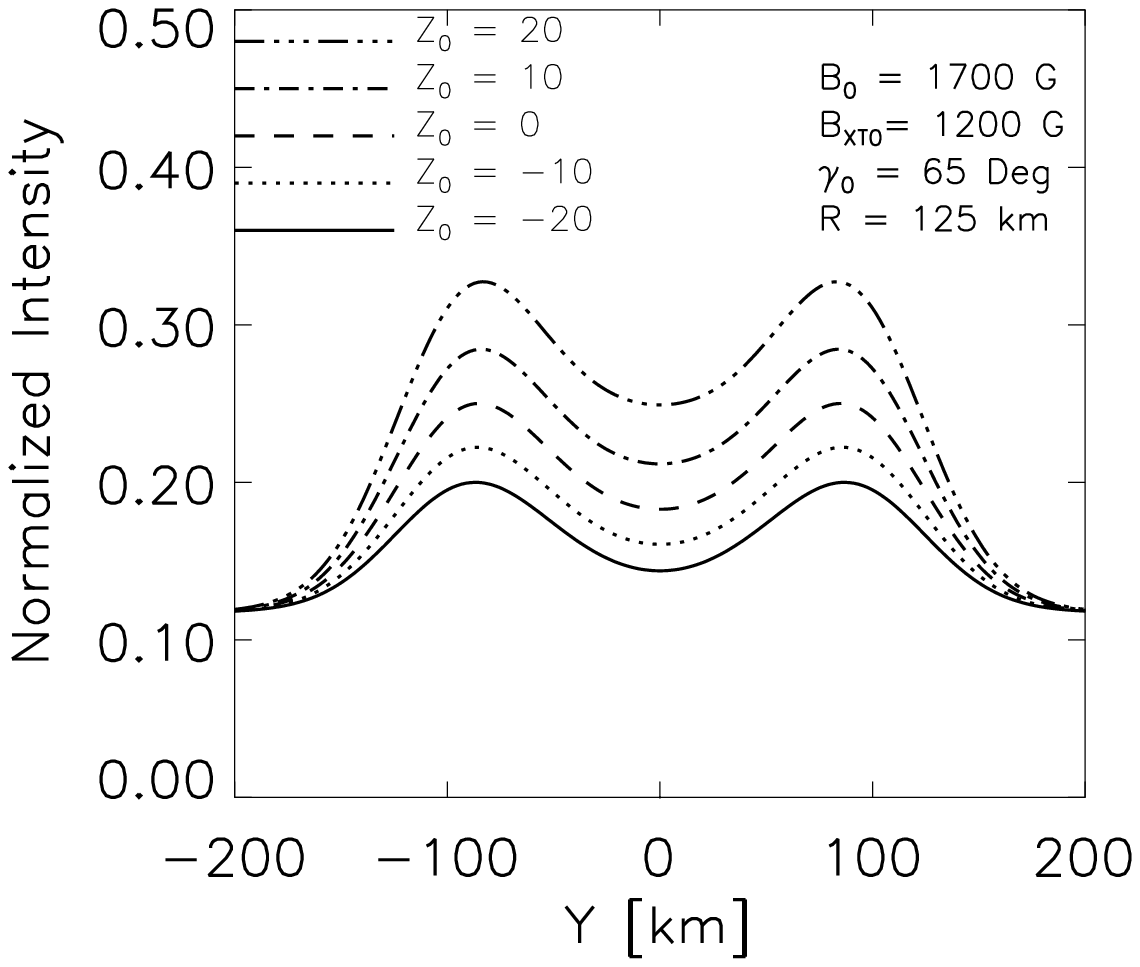} \\
\includegraphics[angle=0,width=7cm]{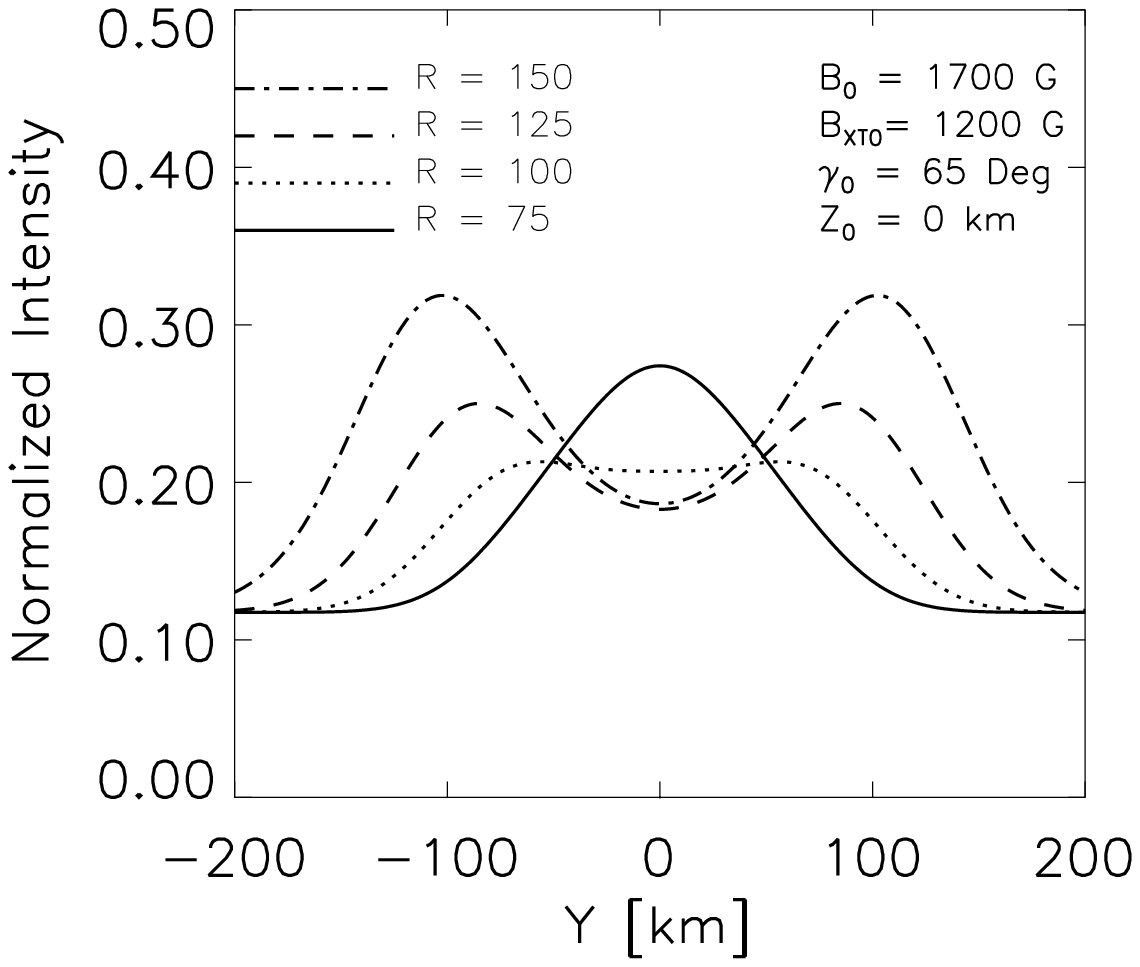} \\
\includegraphics[angle=0,width=7cm]{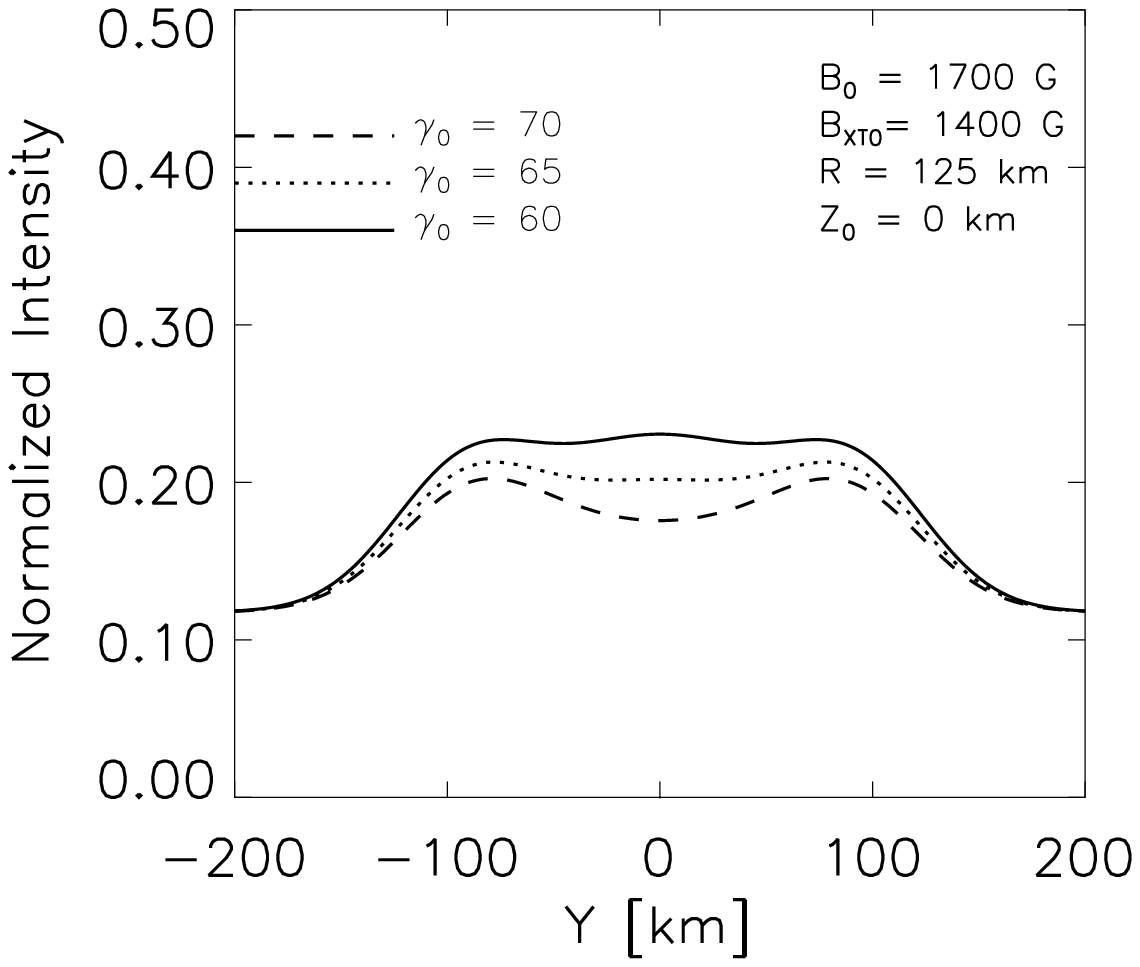}
\end{center}
\caption{Variation of the observed continuum intensity 
at a reference wavelength of 5000 \AA~ along the cross section 
of the flux tube. {\it Top panel}: all parameters are kept constant (see 
text for details) except for the position  of the flux tube axis with 
respect the continuum level in the external atmosphere: $z_0$. {\it Middle panel}: 
same as before but now the radius of the flux tube is the only parameter 
that changes. {\it Bottom panel}: same as before but now what varies is 
the inclination of the external field $\gamma_0$.}
\end{figure}

\begin{figure}
\begin{center}
\includegraphics[angle=0,width=8.5cm]{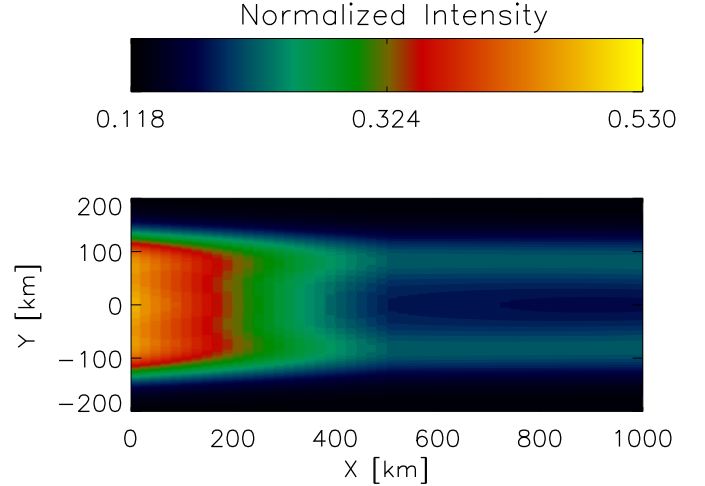}
\end{center}
\caption{Simulated intensity map of a penumbral filament. $\gamma_0$
and $z_0$ change (along the $x$-direction) from 60$^{\circ}$ to 70$^{\circ}$
and from 50 to 0 km. The remaining parameters are: $B_0=1700$ Gauss, 
$B_{tx0}=1300$ Gauss and $R=125$ km.}
\end{figure}

%%%%%%%%%%%%%%%%%%%%%%%%%%%%%%%%%%%
\subsection{Results for $\alpha \ne 0$}%
%%%%%%%%%%%%%%%%%%%%%%%%%%%%%%%%%%%

For $\alpha \ne 0$, functions such as $\mathcal{F}(\theta)=\sin\theta$ or 
$\mathcal{F}(\theta)=\cos\theta$ lead to singular $B_{rt}(r,\theta)$ functions.
(see Eq.~24). However $\mathcal{F}(\theta)=\cos^2\theta$, $\mathcal{F}(\theta)=\sin^2\theta$
and $\mathcal{F}(\theta)=1$ are plausible.

We have repeated all the calculations presented in Sect.~6 for the case
$\alpha \ne 0$ and $\mathcal{F}(\theta)=\cos^2\theta$. In this case,
$\alpha$ is constrained between $\simeq [-10^6,10^6]$ to avoid
singularity problems in Eq.~24 and large density perturbations in Eq.~18.
Results show that the maximum temperature inside the flux tube increases for $\alpha>0$, but 
the opposite occurs for $\alpha<0$ (see Fig.~6). This expands the allowed
ranges in Fig.~3 for $\alpha < 0$ as compared to $\alpha=0$, and vice-versa for $\alpha > 0$.

\begin{figure}
\begin{center}
\includegraphics[angle=0,width=8cm]{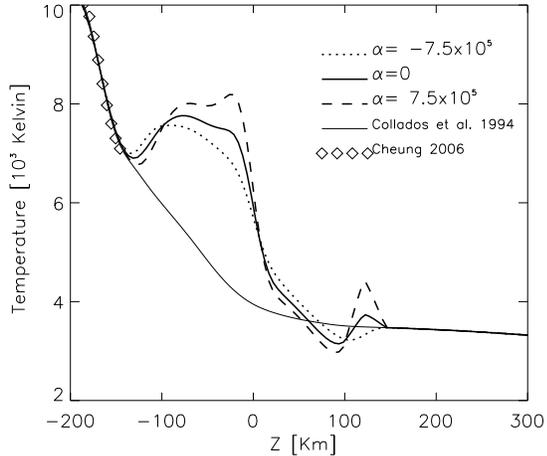}
\end{center}
\caption{Vertical stratification of the temperature along the center of the flux tube: $T(0,z)$
for different values of $\alpha$. Dotted line: $\alpha=-7.5\times10^5$; solid line: $\alpha=0$;
dashed line: $\alpha=7.5\times10^5$. For this example we have adopted: $B_0=1700$ Gauss, $B_{xt0}=1200$
Gauss, $\gamma_0=65^{\circ}$, $R=125$ km, $Z_0=0$ km. Note that $\alpha<0$ produces
a smoother temperature stratification, with smaller maximum temperatures.
The solid thin line corresponds to the surrounding umbral atmosphere (Collados et al. (1994). 
Diamonds represent the extrapolation towards deep layers starting at $z \sim -150$ km (see Sect.~5).}
\end{figure}

%%%%%%%%%%%%%%%%%%%%%%%%%%%%%%%%%%%%%%%%%%%%%%%%
\section{Critical comparison with observations}%
%%%%%%%%%%%%%%%%%%%%%%%%%%%%%%%%%%%%%%%%%%%%%%%%

Although the main intent of this paper is to describe equilibrium
solutions of a horizontal flux tube embedded in a magnetic
surrounding atmosphere, it is unavoidable to make comparisons 
between the model predictions and observations. 
In previous sections we have mentioned cases where that
comparison is positive. In this section we turn our attention to
aspects where our model faces difficulties.

\subsection{Vertical gradients in the field inclination}

In order to reproduce the Net Circular Polarization (NCP)
observed in sunspots, strong vertical gradients in the field
inclination (about 40$^{\circ}$ in about 100 km: S\'anchez
Almeida \& Lites 1992) are required. In papers II and III
we used a \emph{simple} (the flux tube cross section
was assumed to be square) flux tube model where those large 
gradients in the inclination were produced at the lower and upper 
tube boundaries. In this manner, we successfully reproduced
 the observed NCP and the full polarization profiles 
(see also Schlichenmaier et al. 2002; M\"uller et al. 2002 and 
M\"uller et al. 2006).

A closer look to \emph{more realistic} (circular cross section) 
flux tube models highlights a lack of strong enough vertical 
gradients: see Fig.~10. Since $B_z = 0$ on top of the flux tube,
the inclination of the external field there is still 90$^{\circ}$. 
A 40$^{\circ}$ change in the inclination along the vertical direction
can only be attained if we compare the field inclination inside the flux tube
with the external one 150 km above the upper boundary
(compare two vertical dotted lines in Fig.~10; middle panel). 
However, at this point we start moving outside the region in which 
the spectral lines are formed.

A possible solution to the problem is to reduce the radius of the flux tube.
In this case there would be enough vertical length to accommodate the flux tube
and at the same time to let the external field spread upwards long enough to
see significant changes in the field inclination. A second
possibility is to relax the hypothesis of circular cross sections. A
boundary layer with a cusp shape (Sch\"ussler \& V\"ogler 2006; Scharmer \& 
Spruit 2006) allows $B_z > 0$ on top of the flux tube. This of course makes 
the external field more vertical there and increases the $\Delta \gamma$ across
the upper boundary.

\begin{figure}
\begin{center}
\includegraphics[angle=0,width=8cm]{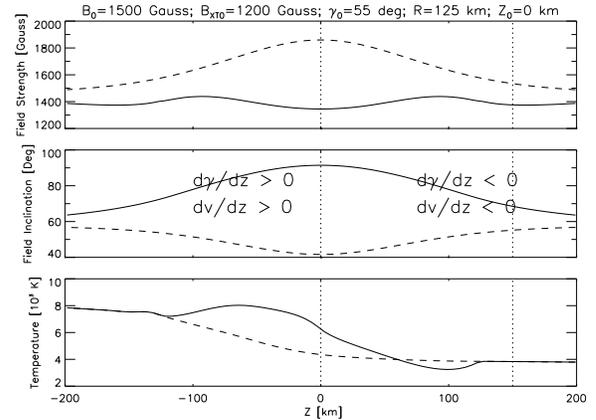}
\end{center}
\caption{Example of the average properties for the flux tube (solid) and surrounding
atmosphere (dashed). From top to bottom: magnetic field strength, magnetic field inclination
and temperature. Note the discrepancies and similarities with the model for embedded flux tubes
used in Borrero et al. (2005; Fig.~4). In the middle panel we indicate the sign of $d\gamma/dz$
and $dv/dz$. The sign of the velocity gradient is calculated using Eq.~3: the flux tube
carries the Evershed flow whereas the surrounding is at rest.}
\end{figure}

\subsection{Lower flux tube boundary}

Perhaps the most serious problem that our model faces is the location of the lower boundary
of the flux tube. According to our calculations and examples shown
here (see Figs.~2 and 6), the $\tau_5=1$ level is always formed inside the flux tube. This
is a consequence of the high temperatures and large opacities present in the lower half of the
flux tube. If our model is correct, this would indicate that the lower boundary
of the flux tube would be invisible to spectropolarimetric observations.
However, such observations seem to indicate (Fig.~7 in Borrero et al. 2005;
see also Fig.~6 in Borrero et al. 2006a) that the lower boundary is visible.

Furthermore, the impossibility of observing the flux tube lower
boundary would make it difficult to explain (without calling upon horizontal gradients) 
the observed change in the sign of $d\gamma/dz$ and $dv/dz$ throughout the penumbra 
(see Fig.~5 in Borrero  et al. 2004). Indeed, this observational result is 
what led Bellot Rubio et al. (2002) and Mathew et al. (2003) 
to first postulate the existence of flux tubes with two defined boundaries. 
Let us keep in mind that, in the flux tube model 
(Fig.~10; middle panel), we have $d\gamma/dz>0$ and $dv/dz>0$ in the lower half of 
the flux tube, but $d\gamma/dz<0$ and $dv/dz<0$ in the upper half. 
It appears now that realistic flux tube models have the region where $d\gamma/dz>0$
and $dv/dz>0$ beyond the line forming region.

%%%%%%%%%%%%%%%%%%%%%%
\section{Conclusions}%
%%%%%%%%%%%%%%%%%%%%%%

We have studied the equilibrium of horizontal magnetic flux tubes 
(with circular cross sections) embedded in an external atmosphere 
with a more vertical magnetic field (assumed to be potential). 
Our study differs from previous investigations in the fact that 
the flux tubes are not necessarily {\it thin}. We have found that, 
for typical penumbral conditions, the equilibrium condition is hardly
satisifed if the flux tube magnetic field is aligned with its axis. 
This situation is alleviated if the tube's magnetic field possesses 
a transverse component.  

We have obtained the thermodynamical configuration resulting
from different possibilities for the transverse magnetic field inside the flux 
tube. We have found that the thermodynamical structure of such flux tubes 
reproduces the dark core penumbral filaments and penumbral grains observed
at high spatial resolution. This happens as a consequence of the high densities
and low temperatures found on the upper part of the flux tube (right above
the continuum $\tau=1$ level formation height).

The overall magnetic configuration explains many (but not all, see Sect. 7),
results from spectropolarimetric observations of sunspot penumbrae. It also explains 
recent observations that find the Evershed effect mainly concentrated in the dark
lanes of penumbral filaments (Bellot Rubio et al. 2005; Rimmele \&
Marino 2006; Langhans et al. 2007). Our Fig.~2 shows that the Evershed
effect appears predominantly at the flux tube's core.
This is due to a larger opacity at the tube's center (where the dark core 
originates) as compared to its sides. Our model also predicts that, 
although weaker, flows should also appear on the brights sides on the 
filaments. This is in agreement with the findings of Bellot Rubio et al. 2005.

Questions such as the stability and temporal evolution of the proposed
configurations are beyond the scope of this work and need to be 
addressed by other means. In particular our approach neglects the energy equation. 
It is important also to note that the presence of high density
plasma region immediately above a low density region makes the model
prone to Rayleigh-Taylor instabilities. It is unclear how the horizontal
field inside the flux tube might help to stabilize the system (Jun et al. 1995).

%%%%%%%%%%%%%%%%%%%
\begin{acknowledgements}
Sami Solanki, Manfred Sch\"ussler, Matthias Rempel, Luis Bellot,
Rolf Schlichenmaier and Goran Scharmer are gratefully acknowledged
for continuous and stimulating discussions on the topic. Thanks
to Hectos Socas-Navarro for carefully reading the manuscript. This work
was completed during a visit to Katlenburg-Lindau with the
finalcial support and sponsorship of the Max Planck Institut 
f\"ur Sonnensystemforschung. This research has made use of NASA's Astrophysics 
Data System.
\end{acknowledgements}
%%%%%%%%%%%%%%%%%%%

%%%%%%%%%%%%%%%%%%%

\end{document}